\documentclass{svproc}
\usepackage[a4paper,margin=1.2in]{geometry} 
\usepackage{cite}
\usepackage{graphicx}
\usepackage{url}

\begin{document}
\mainmatter 

\title{Multi-Agent Architecture in Distributed Environment Control Systems: vision, challenges, and opportunities}

\author{Natasha Astudillo  \and Fernando Koch}
\tocauthor{Natasha Astudillo and Fernando Koch}
\institute{Florida Atlantic University, Boca Raton, FL,\\
\email{\{nastudillo2024,kochf\}@fau.edu}}

\maketitle  
\begin{abstract}
The increasing demand for energy-efficient solutions in large-scale infrastructure, particularly data centers, requires advanced control strategies to optimize environmental management systems. We propose a multi-agent architecture for distributed control of air-cooled chiller systems in data centers. Our vision employs autonomous agents to monitor and regulate local operational parameters and optimize system-wide efficiency. We demonstrate how this approach improves the responsiveness, operational robustness, and energy efficiency of the system, contributing to the broader goal of sustainable infrastructure management. 
\end{abstract}

\section{Introduction}
\label{sec:introduction}


The increasing emphasis on sustainability within the built environment has underscored the need for advanced control strategies in smart buildings. In particular, environmental control systems, such as HVAC, play a critical role, as they are responsible for approximately 50-60\% of energy consumption and 40-60\% of CO\textsubscript{2} emissions \cite{saidur2009a, hidalgo2008a} There is a possibility of implementing distributed control in scenarios featuring numerous devices distributed across multiple structures, by aggregating sensor data from various units and promoting sophisticated and distributed decision-making processes. Recent studies have explored the application of AI-based distributed control solutions, which leverage multi-agent architectures to coordinate and optimize the operation of these systems \cite{fuentes-fernandez2009a}. Such architectures not only enable enhanced system responsiveness and efficiency, but also open new avenues for research in addressing scalability, robustness, and interoperability challenges in complex distributed environments.


We are researching and developing on the application of modern multi-agent architectures to design a distributed control system for scenarios with multiple devices and multiple sites in an same region. We focus on data centers that operate fleets of air-cooled chillers in data center scenarios, typically managing between 30 to 100 devices per building and spanning multiple buildings per site. Each facility presents unique configurations and operational conditions, compounded by extremely strict security and proprietary requirements that require localized, resilient solutions. We are exploring methods for implementing distributed autonomous agents to facilitate real-time monitoring and adaptation, fault tolerance, scalability, and security compliance. 

Our research contributes to the state-of-the-art by providing:

\begin{itemize} 

\item A novel distributed control framework that leverages multi-agent architectures for efficient energy management in data centers. 

\item An in-depth analysis of real-time monitoring, adaptation, and fault-tolerant mechanisms in complex, multi-site environments. 

\item Insights into the integration of security and proprietary constraints into decentralized control systems. 

\item Empirical results that validate the scalability and robustness of our approach under various operational conditions.

\end{itemize}


\section{Proposal}
\label{sec:proposal}

\begin{figure}[h!]
    \centering
    \includegraphics[width=0.8\textwidth]{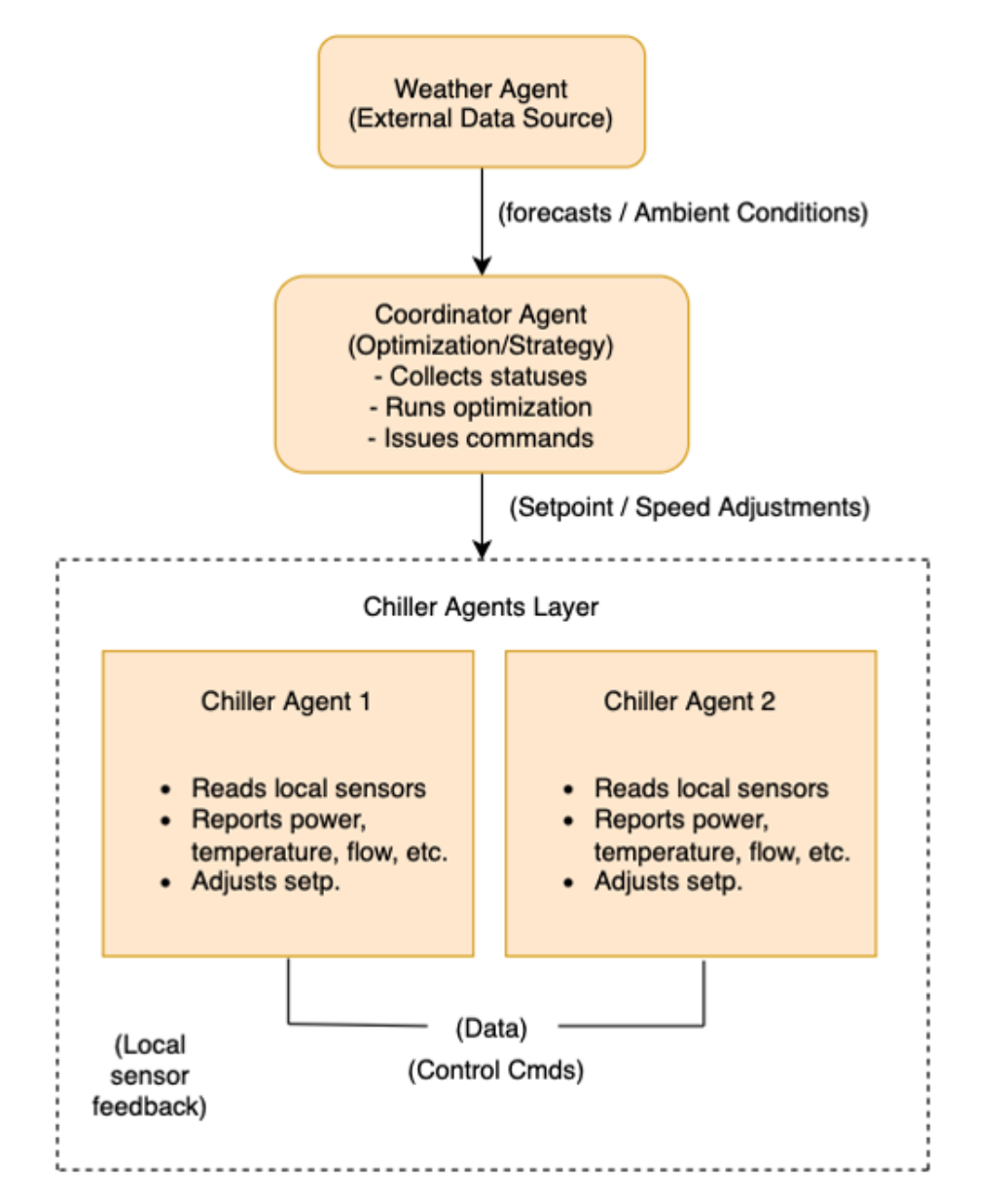}
    \caption{Multi-Agent Systems in Distributed Control Systems}
    \label{fig:sol-arch}
\end{figure}

Recent research highlights the benefits of multi-agent collaboration for energy-efficient building systems \cite{labeodan2015a, qiao2006a}. Solutions built on these architectures can be implemented on premises, aligning with stringent data center security policies while enhancing the overall resilience of the cooling system \cite{yu2022a}. Multi-agent systems (MAS)  minimizes reliance on centralized infrastructure while improving adaptability and efficiency by distributing intelligence across localized agents. Key operational advantages include \cite{koch2005a}:

\begin{itemize} 

\item \textbf{Situatedness:}  agents are situated in an environment which they can influence and be influenced by. 

\item \textbf{Openness:}  agents are able to accommodate changes in the system structure as when, for example, new components enter the system or existing components leave.

\item \textbf{Locality in control:} agents can operate autonomously based on local policies. In mobile services, this may be necessary to ensure robustness of the service.

\item \textbf{Locality in interactions:} agents are able to interact with other components in local geographical or logical neighborhoods.

\item \textbf{Enhanced Security:} Reducing data transmission to external servers mitigates cybersecurity risks.

\item \textbf{Scalability:} Distributed decision-making allows the system to expand dynamically without single points of failure.

\item \textbf{Adaptive Coordination:} Each chiller auto-adapts its setpoints while synchronizing with neighboring units to optimize overall performance.

\end{itemize}

Based on these principles, we propose an integrated solution architecture that incorporates MAS-based distributed control functionalities tailored for complex environmental control systems. The architecture, depicted in Figure \ref{fig:sol-arch}, comprises the following key elements:

\begin{enumerate} 

\item \textbf{Sensors and IoT Devices}: a distributed network of sensors continuously collects high resolution data (e.g. temperature, humidity, occupancy) from various zones within the building.; this granular data collection is essential to capture the unique environmental conditions of each zone, thereby enabling adaptive control strategies.

\item \textbf{Local Deep Reinforcement Learning (RL) Agents:} deployed on edge devices within each building zone, these agents utilize locally collected sensor data to make immediate, context-specific control decisions for HVAC operations; ocal deployment minimizes communication delays and allows each agent to optimize energy consumption based on the specific conditions of its zone. 

\item \textbf{Coordination Layer (Multi-Agent Network):} facilitates robust communication and data exchange between local RL agents, ensuring that their individual actions are harmonized; the coordination layer prevents conflicting control actions between different zones, thus maintaining overall system stability by enabling distributed interactions.

\item \textbf{Central Aggregator:} collects and integrates data from the coordination layer to provide a comprehensive overview of HVAC performance throughout the building; This element ensures that local optimizations contribute to overall energy management and sustainability goals by balancing individual zone performance with global objectives. 

\item \textbf{Cloud Analytics and Re-Training Platform:} processes aggregated data to identify long-term trends, facilitating a deeper understanding of building usage patterns; it supports periodic re-training of the RL models, ensuring that the system adapts to evolving operational conditions and maintains high levels of performance. 

\end{enumerate}

This integrated MAS-based distributed control architecture is designed to address the complex operational challenges inherent in modern data centers and smart building environments. The proposed solution aims to enhance energy efficiency, resilience, and security while simultaneously meeting the stringent requirements of contemporary facility management by combining localized decision-making with coordinated and system-wide oversight.

\section{Use Cases}
\label{sec:use-cases}

Our experiments demonstrate that intelligent redistribution optimizes cooling efficiency without overloading a single unit, leading to an estimated 5-15\% improvement in energy efficiency while maintaining thermal comfort. This use case occurs during periods of high electricity demand, such as hot summer afternoons, when energy costs can surge because of increased cooling requirements. A Coordinator Agent within the multi-agent system (MAS) dynamically optimizes chiller operations to mitigate excessive power consumption. Instead of uniformly increasing cooling loads across all units, the system strategically adjusts operations:

\begin{itemize}
\item Chillers near the building perimeter, where cooler external air intake is available, are directed to operate at higher efficiency.
\item Chillers in hotter zones reduce their load to balance the overall system.
\end{itemize}

In addition, preliminary experiments demonstrate that proactive workload distribution extends the lifespan of the equipment by up to 30\%, reducing unplanned downtime and maintenance costs. Over time, chillers wear and tear, requiring periodic maintenance to maintain optimal performance. MAS enhances adaptive scheduling by continuously monitoring runtime thresholds and dynamically redistributing workloads when necessary. Instead of running all chillers at equal load, the system strategically shifts tasks:

\begin{itemize} 
\item When specific chillers approach their recommended runtime limits, MAS shifts cooling tasks to recently maintained units.
\item This prevents overloading older equipment, reducing mechanical strain and lowering the likelihood of sudden failures.
\end{itemize}

We also demonstrate that keeping all decision-making and learning processes on-premises eliminates cybersecurity risks associated with data transmission, ensuring full compliance with security policies. Data privacy is a critical concern for data centers, especially when optimizing cooling efficiency using AI-based models. MAS enables secure local learning, allowing each site to develop optimization strategies without exposing sensitive operational data to external servers. Instead of relying on cloud-based learning, the system ensures:

\begin{itemize} 
\item Each facility trains its own RL-based optimization models using localized sensor data.
\item All decision-making remains on-premises, eliminating risks associated with external data transmission.
\end{itemize}

Finally, our research explores dynamic chiller adjustments based on real-time weather variations to reduce unnecessary energy consumption and improve system stability. The environmental conditions can change unexpectedly, affecting the efficiency of the cooling system. MAS improves resilience by incorporating a Weather Agent that detects sudden temperature fluctuations and adjusts operations accordingly. Instead of relying on static cooling schedules, the system proactively modifies operations:

\begin{itemize} 
\item If an unanticipated nighttime temperature drop is detected, the coordinator agent adjusts the chiller speeds to prevent unnecessary cooling.
\item During heatwaves, MAS redistributes cooling loads between zones to prevent overheating and improve energy efficiency.
\end{itemize}

\begin{table}[h]
    \centering
    \begin{tabular}{|l|l|l|}
        \hline
        \textbf{Aspect} & \textbf{Standalone (Local Control Only)} & \textbf{MAS (Proposed)} \\
        \hline
        Security & 100\% local but no synergy & 100\% on-premises, min. comms overhead \\
        \hline
        Energy Savings & Baseline (0\%) & 5--20\% lower consumption \\
        \hline
        Fault Tolerance & Reactive/manual response & 30--40\% faster anomaly handling \\
        \hline
        Scalability & Single BMS bottleneck & Distributed, coordinator-based \\
        \hline
        Maintenance Efficiency & Traditional schedules & Up to 30\% improvement in scheduling \\
        \hline
    \end{tabular}
    \caption{Comparative analysis}
    \label{tab:comparison}
\end{table}

These results are summarized in Table \ref{tab:comparison}.

\section{Challenges and Opportunities}
\label{sec:challenges}

The integration of a Multi-Agent System (MAS)-based distributed control architecture for energy-efficient building systems presents several challenges, including:

\begin{itemize}
    
    \item \textbf{Inter-Agent Coordination Complexity:} Ensuring seamless collaboration between autonomous agents can be challenging, especially in large-scale implementations where communication delays and decision conflicts can arise.
    
    \item \textbf{Computational and Hardware Constraints:} Deploying local deep reinforcement learning (RL) agents on edge devices requires computational efficiency, as resource limitations can impact real-time processing and decision-making.
    
    \item \textbf{Security and Privacy Concerns:} Although local decision-making reduces external data transmission, vulnerabilities within the local network may still expose critical control systems to cyber threats.
    
    \item \textbf{Scalability and Interoperability:} Ensuring that the MAS framework can scale effectively while remaining compatible with heterogeneous IoT devices, sensors, and existing building management systems is a key challenge.
    
    \item \textbf{Training Data and Adaptability:} real-world environmental changes may introduce non-stationary, requiring adaptive mechanisms to prevent performance degradation.

\end{itemize}

We seek to address these challenges through a combination of advanced control strategies, system design optimizations, and adaptive learning mechanisms.

\begin{itemize}
    \item \textbf{Enhanced Inter-Agent Coordination:} we are exploring hierarchical multi-agent coordination to mitigate communication delays and decision conflicts. A coordination layer facilitates synchronized interactions, ensuring that agents optimize their decisions collaboratively while minimizing disruptions.
    \item \textbf{Efficient Computational Strategies:} we are researching how to optimize reinforcement learning models for edge deployment by making use of dynamic Generative AI Pipelines and, eventually, integrating Small Generative Models to overcome hardware and computational constraints.
    \item \textbf{Robust Security and Privacy Mechanisms:} we are designing our MAS framework to integrate zero-trust security models, encrypted local communication, and anomaly detection algorithms to safeguard critical infrastructure from cyber threats. 
    \item \textbf{Scalability and Interoperability Solutions:} we are designing the MAS architecture with standardized communication protocols (e.g., MQTT, OPC-UA) and modular software interfaces, to ensure seamless expansion and compatibility, 
    \item \textbf{Adaptive Learning for Real-World Variability:}  we are implementing techniques for Hybrid Ai and  continuous learning frameworks where MAS agents periodically update their models using online reinforcement learning to address the challenges of non-stationary environments,
\end{itemize}

We highlight several key opportunities that make this integration very promising. First, the adoption of \textbf{distributed decision-making} enables dynamic and real-time optimization of HVAC operations, reducing energy consumption and operational costs while maintaining optimal occupant comfort. By continuously adapting to changing conditions, MAS-based control strategies can lead to substantial energy savings and improved system performance. Moreover,  \textbf{localized control capabilities of MAS} reduce dependence on centralized infrastructure, ensuring robustness against network failures and system failures. Moreover, adaptive control mechanisms driven by \textbf{ MAS} can dynamically adjust to environmental fluctuations and occupancy patterns, ensuring optimal energy efficiency in varying conditions.  The proposed architecture can be extended to interact with smart grids, enabling demand-response strategies and facilitating the seamless integration of renewable energy sources.  

These opportunities underscore the transformative potential of MAS-based distributed control systems, making them a viable solution for future-proofing energy-efficient buildings while addressing both economic and environmental sustainability goals.

\section{Conclusions}
\label{sec:conclusions}

The integration of MAS into distributed control systems represents a paradigm shift in energy-efficient building management. As advancements in AI, IoT, and decentralized intelligence continue to evolve, MAS-based architectures have the potential to drive next-generation sustainable infrastructure, ensuring optimal energy utilization, resilience, and future-proofed smart building solutions.

Our preliminary findings demonstrate that MAS enhances energy efficiency, resilience, scalability, and security while maintaining compliance with stringent operational constraints. The preliminary analysis identified key operational advantages of MAS, including improved fault tolerance, adaptive scheduling, and decentralized decision-making. Experimental results indicate that MAS-driven control can lead to 5--20\% energy savings, 30--40\% faster anomaly detection, and up to 30\% improvement in predictive maintenance scheduling. Several areas for further research and advancements remain:

\begin{itemize}
    \item \textbf{Advanced Generative AI for Enhanced Learning:} Future work could explore the integration of Generative AI Pipelines and Small Generative Models to improve reinforcement learning efficiency and enable faster adaptation to dynamic environments.
    
    \item \textbf{Hybrid AI for Enhanced Adaptability:} Investigating Hybrid AI approaches, combining symbolic reasoning with deep learning, could help MAS agents better interpret environmental changes and make more explainable and context-aware decisions.
    
    \item \textbf{Interoperability with Emerging IoT Standards:} As IoT ecosystems evolve, ensuring seamless interoperability with next-generation smart building technologies and industry standards (e.g., Matter, OPC-UA, MQTT) will be critical for real-world deployments.
    
    \item \textbf{Decentralized Trust and Security Mechanisms:} Further research is needed to strengthen privacy-preserving computation methods, such as federated learning and zero-trust security architectures, ensuring robust protection against cyberthreats while maintaining distributed autonomy.
    
    \item \textbf{Integration with Smart Grids and Renewable Energy Systems:} Extending MAS architectures to interact with smart grids and renewable energy sources can enhance demand-response capabilities, promoting sustainable energy consumption in future buildings.
\end{itemize}

%
%
\bibliographystyle{IEEEtran} 
\bibliography{references} 

\end{document}